\documentclass[aps,prb,preprint,superscriptaddress,amsmath,amssymb]{revtex4}
\usepackage{graphicx}
\usepackage{dcolumn}
\usepackage{bm}
\begin{document}
\title{Density functional theory study of the nematic-isotropic transition 
in an hybrid cell}
\author{I. Rodr\'{\i}guez-Ponce}
\affiliation{Physik Department, Technische Universit\"at M\"unchen,
James-Franck-Strasse, D-85747 Munich, Germany}
\author{J. M. Romero-Enrique}
\affiliation{Departamento de F\'{\i}sica At\'omica, Molecular y Nuclear,
Area de F\'{\i}sica Te\'orica, Universidad de Sevilla, Apartado de correos 
1065, 41080 Sevilla, Spain}
\affiliation{Department of Mathematics, Imperial College 180 Queen's Gate,
London SW7 2BZ, United Kingdom}
\author{L. F. Rull}
\affiliation{Departamento de F\'{\i}sica At\'omica, Molecular y Nuclear,
Area de F\'{\i}sica Te\'orica, Universidad de Sevilla, Apartado de correos 
1065, 41080 Sevilla, Spain}
\begin{abstract}
We have employed the Density Functional Theory formalism to investigate 
the nematic-isotropic capillary transitions of a nematogen confined by walls 
that favor antagonist orientations to the liquid crystal molecules 
(hybrid cell). We analyse the behavior of the capillary transition as a 
function of the fluid-substrate interactions and the pore width. 
In addition to the usual capillary transition between isotropic-like to
nematic-like states, we find that this transition can be suppressed 
when one substrate is wet by the isotropic phase and the other by the nematic
phase. Under this condition the system presents interface-like states which
allow to continuously transform the nematic-like phase to the isotropic-like
phase without undergoing a phase transition. Two different mechanisms for the 
disappearance of the capillary transition are identified. When the director of 
the nematic-like state is homogeneously planar-anchored with respect to the 
substrates, the capillary transition ends up in a critical point. This 
scenario is analogous to the observed in Ising models when confined in slit 
pores with opposing surface fields which have critical wetting transitions. 
When the nematic-like state has a linearly distorted director field, the 
capillary transition continuously transforms in a transition between two 
nematic-like states.  
\end{abstract}
\maketitle
\section{Introduction}
The effect on confinement on simple fluids has been vastly investigated 
in the past. Most of studies are related to the effect of the confinement 
on the vapor-liquid transition, the so-called capillary transition. In 
the case of symmetric walls, where both walls attract or repel the 
molecules with the same strength, the capillary transition in large 
pores is governed by the well-known Kelvin law \cite{Kelvin}.  
In addition, it may appear surface phenomenology that can affect the 
global phase diagram. Often this surface phenomenology is related to 
wetting states found in the single wall cases. The latter phenomenon 
is related to the spreading of drops of vapor (liquid) on the surface 
in coexistence with the bulk liquid (vapor). The presence of these surface 
states can be considered in the description of the phase behavior in the 
confined system. The generalized Kelvin law for small pores (where surface 
phenomenology is more dominant) takes into account these effects so it is 
still possible to determine the vapor-liquid transition for small pores 
\cite{Tarazona86}. However, the predictions of the generalized Kelvin law 
reveal that the wetting phenomena does not affect essentially the phase 
diagram of the confined problem so they are often discarded. An interesting 
system is the \emph{hybrid} cell (also known as asymmetric cell) where one 
substrate or wall is repulsive and the other attractive (competing walls). 
In this case, the wetting properties present in the single wall cases play 
a crucial role in the determination of the critical point in the confined 
problem. Here, given a pore width, the coexistence of two phases can only 
occur for temperatures below $T_W$, which corresponds to the total wetting 
temperature of the single wall cases \cite{Parry,Indekeu}.

These studies have elucidated which mechanisms are involved in the surface 
transitions and how the properties of the substrate can determine the phase 
behavior within the pore. With respect to complex fluids like liquid crystal 
(LC) there is an increasing interest in the understanding of the 
surface phenomena due to its crucial role in the advance of the LC technology. 
For the case of a single wall (semi-infinite problem) and symmetric pores 
favoring an specific orientation on the molecules, theoretical approaches 
\cite{Fonseca,Kocevar,Gama84,Teixeira93,Sen87,Poniewierski00,Seidi97} as 
well as computer simulations \cite{Zannoni02,Zannoni98,Allen99,Cleaver97,
Pelcovits00} reveal that the interaction wall-fluid particle can 
affect strongly the ordering not only close to the surface but also far 
from it. Additionally to finite size effects, novel phenomena often 
appear due to the combined presence of anchoring and orientational wetting 
transitions. The interplay between both surface phenomena and its dependence 
on wall properties have been recently studied by us employing a microscopic 
Density Functional Theory (DFT) approach for a simple model of liquid crystal 
\cite{paper1}. The effect of confinement by symmetric walls on these surface 
transitions was also studied \cite{paper2}. Our findings revealed that as 
in the case of simple fluids, the finite-size effects for not very narrow pores
on the phase diagram are not of qualitative significant importance  
and only the location of the capillary transitions are affected by the 
confinement. However, recent theoretical \cite{Evans00,Heras} and
simulation results \cite{Evans00} for hard-spherocylinder models 
show that for very narrow symmetrical pores the NI capillary transition may 
end up in a critical point. 

In a recent paper \cite{paper3}, we extended the study to the case with
an hybrid cell where the surfaces favor antagonist orientations. This 
geometry has in LCs a particular interest due to the possible technological 
applications. This fact has stimulated an increasing interest by 
experimentalists and theoreticians \cite{Wittenbrood98,Zumer99,Evans00}. 
In particular, different nematic-like states with homogeneous and linearly
distorted director fields were found to coexist under confinement.  
In the present paper, our goal is to complete this study with the 
investigation of the nematic-isotropic (NI) transition within the 
hybrid cell. Quintana {\it et al} investigated a similar problem employing a
Landau-type theory \cite{Robledo98}. They claimed that the NI transition 
either appears unaffected and its temperature $T_{NI}$ remains unchanged or 
the transition disappears. Basically what determined the existence of the 
transition was the value of fluid-wall interaction strength with respect 
to the value for the first-order wetting transition for the semi-infinite 
systems. For strength values below this threshold the transition occurs for 
all wall separations. In contrast, above the threshold, the transition 
disappears for large pore widths, where interfacial-like states appear. 
These states are related to the wetting states by isotropic and nematic phase
that each wall present in the semi-infinite geometry, respectively.

In this paper we address the same problem of the behavior of the
NI capillary transition and its possible disappearance from a more microscopic 
point of view by using a DFT approach. Our paper is organized as follows. 
First we briefly present our liquid crystal model in Section II. In Section 
III we revisit and complete the study of a substrate which favors either 
planar or homeotropic anchoring in contact with a nematogen fluid at its 
isotropic or nematic phase. In particular, the different wetting regimes are 
found. This phenomenology will be a crucial guide to understand the behavior 
of the fluid under confinement. Section IV is devoted to the study of the 
fluid adsorbed in a hybrid cell, paying special attention to the disappearance 
of the NI capillary transitions and the mechanisms involved in it. Finally we 
end up with our conclusions. 

\section{The Model}
The theoretical model is a standard generalized van der Waals theory 
based on a perturbative expansion, using a hard-sphere (HS) fluid
as reference system \cite{Gama84}. Details on the physical basis of the 
model and how to obtain its solutions numerically can be found elsewhere
\cite{anchoring,minimo}. Our starting point is the grand potential functional
per unit system area $A$, $\Omega[\rho]/A$, whose functional minimum with 
respect to the one-particle distribution function, 
$\rho({\bf r},\hat{\bf\Omega})$, which depends on both molecular positions 
${\bf r}$ and orientations $\hat{\bf\Omega}$, gives the equilibrium 
structure of the interface. This function, $\rho({\bf r},\hat{\bf\Omega})
\equiv\rho(z)f(z,\hat{\bf\Omega})$, contains a mass distribution 
$\rho(z)$ and an angular distribution $f(z,\hat{\bf\Omega})$.
These quantities vary locally with the distance from $z=0$ to $z=H$, where 
$H$ is the pore width. The expression for $\Omega[\rho]$, in a mean field 
approximation is,
\begin{eqnarray}
&&\Omega[\rho]=F_{r}[\rho]+\frac{1}{2}\int\!\!\int\!\!\int\!\!\int d{\bf r}
d{\bf r}^{\prime}d\hat{\bf\Omega}d\hat{\bf\Omega}^{\prime}
\rho({\bf r},\hat{\bf\Omega})\rho({\bf r}^{\prime},\hat{\bf\Omega}^{\prime})
\nonumber\\
&&\hspace{0cm}\times v({\bf r}-{\bf r}^{\prime},\hat{\bf\Omega},
\hat{\bf\Omega}^{\prime})-\int\!\!\int d{\bf r}d\hat{\bf\Omega}
\rho({\bf r},\hat{\bf\Omega})[\mu-v_W({\bf r},\hat{\bf\Omega})],
\end{eqnarray}
where $\mu$ is the chemical potential and
\begin{eqnarray}
F_{r}[\rho]=\int d{\bf r}f_{hs}(\rho({\bf r})) +
k_BT\int d{\bf r}\rho({\bf r})\langle\ln(4\pi f(z,\hat{\bf \Omega}))\rangle
\end{eqnarray}
is the reference system free energy. In the above expression
${f}_{hs}(\rho({\bf r}))$ is the hard-sphere free energy
density of a uniform fluid with a density equal to the local density at
${\bf r}$ and $\langle\ldots\rangle$ is an angular average. 
The attractive potential $v$ contains anisotropic (dispersion) forces 
driving the liquid-crystalline behavior of the model material:
\begin{eqnarray}
v({\bf r},\hat{\bf\Omega},\hat{\bf\Omega}^{\prime})=
v_A(r)+v_B(r)P_2(\hat{\bf\Omega}\cdot\hat{\bf\Omega}^{\prime})\\ \nonumber
+v_C(r)\left[P_2(\hat{\bf\Omega}\cdot{\hat{\bf r}})+
P_2(\hat{\bf\Omega}^{\prime}\cdot{\hat{\bf r}})\right],
\label{pot}
\end{eqnarray}
where $\hat{\bf r}={\bf r}/r$, and $r$ is the center-of mass
distance. The functions $v_A(r)$, $v_B(r)$, $v_C(r)$ are the radial 
contribution to the potential as a result of considering dispersion-type 
forces. In this work we choose them to have a simple Yukawa form, i.e. \
$v_i(r)=-\epsilon_i\exp{(-\lambda_i (r-d))}/r$ for $r>d$, 
and $v_i(r)=0$ otherwise, where $d$ is the diameter of a hard sphere. 
We note that a positive value of $\epsilon_C$ favors the molecule to 
orientate parallel to the NI interface and a negative value favors 
the perpendicular orientation to NI interface. This {\it anisotropy} 
in the orientational configuration of the nematic molecules is essential 
to understand the phenomenology in the problem with the walls. 

The walls are modelled via the following potentials:
\begin{equation}
v^1_W(z,\theta)=\begin{cases}
+\infty & z<0\\
-\epsilon_W^1 
e^{-\lambda_W (z-d)}P_2(\cos{\theta})& z>0\\
\end{cases}
\label{homeotropic}
\end{equation}
where $\theta$ is the angle between the molecule axis and the $z$-axis normal
to the substrate. For $\epsilon_W^1>0$, this potential favors the molecules
to align perpendicularly to the surface (homeotropic anchoring). The potential
corresponding to the other substrate is:
\begin{equation}
v^2_W(z,\theta)=\begin{cases}
+\infty & z>H \\
\epsilon_W^2 e^{-\lambda_W (-(H-z)-d)}P_2(\cos{\theta}) & z<H\\
\end{cases}
\label{planar}
\end{equation}
which favors the molecules to align in a plane parallel to the 
surface for $\epsilon_W^2>0$. However, this preferred alignment corresponds
to a random planar anchoring, since there is not a preferred
direction at the $xy$ orientation plane. The parameters $\epsilon_W^1$
and $\epsilon_W^2$ measure the substrate-fluid interaction strength of 
each substrate.

In our study, we will assume that the nematic director 
orientation is embedded in the $xz$ plane. The formalism permits to define 
the orientation distribution of the molecules referred to a laboratory
reference system, described by three order parameters,
\begin{eqnarray}
\eta(z)=\int d\phi \sin \theta d\theta f(z,\hat \Omega)P_2(\cos \theta) \\
\nu(z)=\int d\phi\sin \theta d\theta f(z,\hat \Omega)\sin2 \theta\cos \phi \\
\sigma(z)=\int d\phi\sin \theta d\theta f(z, \hat \Omega)\sin^2 \theta \cos2
\phi
\end{eqnarray}
The order also can be referred to an intrinsic reference system by three 
quantities: $\psi$ the {\it tilt} angle defined like the angle formed by the 
director with the $z$ axis, $U$ (uniaxial order parameter) the amount of 
order along the director and $B$ (biaxial order parameter) which measures 
the amount of order along the perpendicular directions of the director. 
The nematic ordering is essentially uniaxial ($B=0$) far from the substrate. 
The biaxiality induced by the surfaces was already discussed in a previous 
paper \cite{paper3}.

These intrinsic order parameters can be obtained in terms of 
the set $\{\eta,\sigma,\nu\}$ by the following expressions,
\begin{eqnarray}
\tan \psi(z)=\frac{\nu(z)}{\eta(z)-\frac{\sigma(z)}{2}+\sqrt{\left[\eta(z)-
\frac{\sigma(z)}{2}\right]^2+\nu(z)^2}}  \\
U(z)=\frac{\eta(z)}{4}+\frac{3}{8}\sigma(z)+\frac{3}{4}\sqrt{\left[\eta(z)-
\frac{\sigma(z)}{2}\right]^2+\nu(z)^2}  \\
B(z)=
\frac{\eta(z)}{2}+\frac{3}{4}\sigma(z)-\frac{1}{2}\sqrt{\left[\eta(z)-
\frac{\sigma(z)}{2}\right]^2+\nu(z)^2}
\end{eqnarray}
It is convenient to define the average density and order parameter for 
the confined system as: 
\begin{eqnarray}
\bar \rho=\frac{1}{H}\int_0^H  \rho(z)dz\\
\bar U=\frac{1}{H}\int_0^H  U(z)dz
\end{eqnarray}
As we will also make reference to the semi-infinite problems, the 
adsorption in the order parameter is defined as:
\begin{eqnarray}
\Gamma=\int_0^\infty (U-U_b) dz
\end{eqnarray}
where $U_b$ is the value of the intrinsic nematic order parameter of the 
bulk phase.

Numerical values for the potential parameters were taken as
$\epsilon_A=1$ (which sets the temperature scale), $\epsilon_B/\epsilon_A =
0.847$ and $\epsilon_C/\epsilon_A=0.75$. The range parameters 
$\lambda_i$ are set, in units of $d$ (throughout we choose 
this unit to set the length scale), to $\lambda_i=2,4,1.75$, $i=A,B,C$ 
respectively, and $\lambda_W=1$. In our calculations the temperature is 
fixed at the value $T^*\equiv k_B T/\epsilon_A=0.57$, that corresponds to 
a typical situation in which vapor, isotropic and nematic phases can be 
observed. At this temperature, the isotropic-nematic coexistence occurs for
a reduced chemical potential $\beta\mu\equiv \mu/k_B T =-3.918$.

\section{The semi-infinite case.}

In order to better understanding the behavior of the fluid inside the
pore, we have first considered the semi-infinite cases of the fluid
in presence of a single substrate that either favors homeotropic or planar
alignment, i.e. the substrate-fluid interaction potentials are given 
by Eq. (\ref{homeotropic}) with $\epsilon_W^1>0$ or $\epsilon_W^1<0$, 
respectively. We are interested in the wetting behavior, so the system
will be at bulk nematic-isotropic coexistence, unless stated otherwise.
We have considered both situations in which the bulk fluid is either 
isotropic or nematic. 

First we consider the case of a nematic bulk phase in presence of the
substrate. This situation was studied for substrates that favor homeotropic 
anchoring in Ref. \onlinecite{paper1}. We extend this analysis to substrates 
which favor planar alignment. For both negative and small positive values of 
$\epsilon_W^1$ the fluid far from the substrate orientates parallel to the 
substrate. For the case in which $\epsilon_W^1<0$, biaxiality is developed 
in a microscopic layer close to the substrate, 
leading to the decay of the orientational order in that area. If the substrate 
favors a homeotropic alignment, we observe a microscopic isotropic-like layer 
close to the substrate, corresponding to a partial wetting situation of the 
substrate-nematic interface by the isotropic phase. As $\epsilon_W^1$ is 
increased, the isotropic layer close to the substrate becomes macroscopic 
as the isotropic phase completely wets the nematic-substrate interface at 
$\epsilon_W^1=\epsilon_W^{w1}$. Our calculations showed that the wetting 
transition is continuous. The orientation far from the substrate remains 
planar in the complete wetting case. By further increasing $\epsilon_W^1$, 
an anchoring transition occurs for $\epsilon_W^1=\epsilon_W^d$, as the 
nematic orientation changes from planar to homeotropic far from the substrate, 
and the wetting becomes partial again since the isotropic-nematic interface 
favors a planar anchoring in our model. The anchoring transition is first 
order and continues in the single nematic phase region. However, the complete 
wetting with planar orientation state remains metastable up to $\epsilon_W^m$. 
This fact will be important for the discussion of the confined cases.
For the model parameters considered in this paper, $\epsilon_W^{w1}/\epsilon_A=
0.470$, $\epsilon_W^d/\epsilon_A=0.527$ and $\epsilon_W^m/\epsilon_A=0.72$.

We turn to the case in which there is an isotropic phase in bulk.
For small values of $|\epsilon_W^1|$, the inhomogeneities of the density
and order parameter profiles are restricted to a microscopic layer close
to the substrate. Although the fluid is isotropic far from the substrate, 
the anisotropic character of the substrate-fluid potential induces a random 
planar ordering near the substrate, i.e. the particles orientate parallel to 
the substrate but without any preferred direction in that orientation plane
($\eta<0$, $\sigma=\nu=0$). 
Such a condition can be seen as an extreme biaxial case, where $B=2U$, i.e. 
the highest eigenvalue of the orientational order parameter tensor is twofold 
degenerate. The layer width remains microscopically finite as the NI 
coexistence is approached, corresponding to a partial wetting situation. 
For large enough values of $|\epsilon_W^1|$, we observe complete wetting 
situations. For negative values of $\epsilon_W^1$ and at the NI coexistence, 
there is a sudden change at $\epsilon_W^1=\epsilon_W^{w2}$ from the previous 
surface state to a complete wetting state of the isotropic-substrate 
interface by the nematic phase. This corresponds to a first order wetting 
transition. For our model parameters, $\epsilon_W^{w2}/\epsilon_A=-0.359$.
Due to its first order character, there is a prewetting first-order 
transition for $\epsilon_W<\epsilon_W^{w2}$ (see Fig. \ref{fig1}). 
At this transition the particles at the first layer orientates along some 
direction parallel to the substrate, breaking the rotational symmetry of the 
state (see insets of Fig. \ref{fig1} for the order parameter profiles). 
Consequently, the prewetting transition can be seen as a two-dimensional 
isotropic to nematic transition at the fluid layer close to the substrate. 
We will denote $I$ and $I^N$ to the surface states with the random planar 
and nematic layer close to the wall, respectively. For the $I^N$ state, there 
is also biaxiality near the substrate, as occurred when there was a nematic 
fluid in bulk. A similar behavior was found for the Zwanzig model and hard 
spherocylinder fluids \cite{Evans00}, although their ``prewetting'' transition 
is continuous (second order for the Zwanzig model and it is expected to be 
Kosterlitz-Thouless-like for the spherocylinder system). The prewetting
surface nematization transition shifts towards lower chemical potentials as
$\epsilon_W$ is increased, and eventually disappears at a critical point.
%

For large positive values of $\epsilon_W^1$, i.e. $\epsilon_W^1/\epsilon_A>
\epsilon_W^{w3}/\epsilon_A=0.691$, we also found that the nematic
phase completely wets the isotropic-substrate interface. We have checked 
indirectly this by studying the numerically obtained profiles that minimize our
free energy functional. The order parameter profiles that we obtain in our 
numerical minimization are inhomogeneous along the wetting layer 
(see Fig. \ref{fig2}). This fact is due to the mismatch of anchoring 
conditions at the interfaces. We first note that the \emph{intrinsic} order 
parameters are homogeneous along the wetting layer except close to the 
interfaces (see inset in Fig. \ref{fig2}). So the origin of the inhomogeneity 
of the \emph{extrinsic} order parameters is the change of the orientation 
of the nematic director field along the wetting layer. As previously mentioned,
the molecules align parallel to the interface between a nematic and isotropic 
phase. On the other hand, the molecules anchor homeotropically close to the 
substrate due to the strong anchoring coupling between the substrate and the 
fluid. As the nematic layer width diverges, it is possible to fulfill both 
anchoring conditions by slowly varying the nematic director from 
homeotropic to planar anchoring. However, any numerical calculation requires 
minimization in a finite range $R$ for the $z$ coordinate, imposing bulk 
isotropic conditions for $z>R$. In this situation, the nematic director profile
that corresponds to the minimum of the grand-canonical free energy is expected 
to be linear, as it is indeed observed by numerical minimization (see inset of 
Fig. \ref{fig2}). This fact induces a $1/R$ dependence of the apparent 
equilibrium excess grand-canonical free energy (or surface tension $\Sigma$) 
due to the elastic deformation. Such decay is much slower than the usual 
exponential correction when elastic deformations are not involved, and we have 
to consider much larger values of $R$ to obtain reliable values of the true 
equilibrium free energies. As $R$ increases, the apparent surface tension 
decreases and it is always larger than the complete wetting state excess free 
energy, that it is the sum of the surface tension $\Sigma_{wN}$ corresponding 
to the interface between substrate and a nematic phase oriented homeotropically
far away, and the surface tension $\Sigma_{NI}$ between corresponding to the 
nematic-isotropic interface where the nematic phase is planar-anchored
with respect to the interface. The latter surface tension is independent of
the fluid-substrate interactions, and its value is $\beta \Sigma_{NI} d^2
\approx 0.0259$ for the NI coexistence at $T^*=0.57$.
%

Fig. \ref{fig3} summarizes the results of this Section. We represent the 
cosine of the contact angle $\theta$ of a droplet of isotropic phase in 
coexistence with bulk the nematic phase on a planar substrate as a function 
of $\epsilon_W$. The Young equation states that:
\begin{equation}
\cos \theta = \frac{\Sigma_{wN} - \Sigma_{wI}}{\Sigma_{NI}}
\label{young}
\end{equation}
where $\Sigma_{wN}$, $\Sigma_{wI}$ and $\Sigma_{NI}$ are the surface
tensions corresponding to the substrate-nematic, substrate-isotropic and
nematic-isotropic interfaces, respectively. For $\epsilon_W<\epsilon_W^d$ 
(including negative values), the nematic director is parallel to the
substrate far from the substrate. Otherwise, the nematic director is 
homeotropically anchored. The complete wetting region corresponds to $\cos 
\theta=1$. The curve meets the $\cos \theta=1$ line at $\epsilon_W=
\epsilon_W^{w1}$ with no change in slope. This fact is a signature of the
second-order character of the wetting transition. On the contrary, the curve
departs from the $\cos \theta=1$ line at $\epsilon_W=\epsilon_W^d$ with 
non-zero slope, indicating the first-order character of the dewetting 
transition. 
%

Fig. \ref{fig3} also gives information about the wetting properties when there
is an isotropic phase in bulk. By reversing the roles of the nematic and 
isotropic phases in Eq. (\ref{young}), it is straightforward to see that 
the contact angle $\theta'$ of a nematic droplet on a planar substrate
is related to $\theta$ via $\theta'=\pi-\theta$. Consequently, when 
$\cos \theta = -1$ we observe complete wetting by nematic phase of the 
isotropic-substrate interface. This fact is clear for $\epsilon_W < 
\epsilon_W^{w2}<0$. However, for $\epsilon>\epsilon_W^{w3}$ the apparent values
of $\cos \theta$ are less than $-1$. As previously discussed, this is a 
finite-size effect of our numerical calculations and the predicted value is 
$\cos \theta = -1$. Nevertheless, since the nematic layer remains finite as
$\epsilon_W \to \epsilon_W^{w3}$ from below, the localization of the wetting
transition is quite accurate. For both wetting transitions at $\epsilon_W^{w2}$
and $\epsilon_W^{w3}$ there is a change of slope of the $\cos \theta$ curve, 
that it is in agreement with their first-order character. 

\section{The confined case.}

We turn to the case of the fluid confined in a hybrid slit pore, where each 
substrate favors an antagonist anchoring, i.e. homeotropic and planar, 
respectively. Under confinement, it is well known that the bulk transitions
have lower-dimensional counterparts that are known as capillary
transitions. The thermodynamic conditions (temperature, pressure, chemical
potential) for such transitions are usually shifted with respect to the bulk 
coexistence values. Before considering our DFT results, we consider the 
predictions of the Kelvin equation, which gives the leading order for the
shift on chemical potential $\Delta \mu=\mu_{NI}^c-\mu_{NI}^b$
of the capillary nematic-isotropic transition $\mu_{NI}^c$ with respect 
to the bulk value $\mu_{NI}^b$ at large pore widths $H$. Thermodynamic 
considerations \cite{Tarazona86,Poniewierski87,Heras} show that the 
chemical potential shift $\Delta \mu$ of a capillary transition with respect 
to the bulk one between the $\alpha$ and $\beta$ phases at a given temperature 
obeys asymptotically:
\begin{equation}
\Delta \mu \approx \frac{\Sigma_{\alpha \beta}(\cos \theta_1 + \cos\theta_2)}
{(\rho_\beta-\rho_\alpha)H}
\label{kelvin}
\end{equation}
In this expression, $\Sigma_{\alpha \beta}$ is the surface tension for an
$\alpha-\beta$ interface, and $\rho_\alpha$ and $\rho_\beta$ are the bulk 
densities of the $\alpha$ and $\beta$ phases, respectively. Finally,
$\theta_1$ and $\theta_2$ are the contact angles corresponding to the
droplets of $\alpha$ phase that nucleate on each substrate when the
fluid is in bulk in the $\beta$ phase, respectively. In our case, we can 
identify $\alpha$ as the isotropic phase and $\beta$ as the nematic phase.
It is implicitely assumed in Eq. (\ref{kelvin}) that there is a partial 
wetting situation for both substrates in semi-infinite geometries. However, 
if either there is complete wetting only at one substrate or in both but by 
the same phase ($\theta_1=\theta_2=0$ or $\pi$), the effective pore width in 
Eq. (\ref{kelvin}) is reduced to $H-\kappa l$, where $\kappa$ is a constant 
that depends on the range of the wall-fluid interactions and $l$ is the sum 
of the widths of the adsorbed wetting layers. Nevertheless, this modification 
does not change qualitatively the results. Its only effect is to observe
the behavior predicted by the Kelvin equation at larger values of the
pore width $H$ than for a partial wetting situation. 

A different situation arises when both substrates are completely wet by  
\emph{different} phases at coexistence, i.e. $\cos\theta_1\cos\theta_2=-1$.
Studies for symmetrically opposing surface fields Ising systems 
\cite{Parry,Indekeu} and Landau-de Gennes models of liquid crystals 
\cite{Robledo98} show that in this situation there is \emph{no} capillary 
transition. Instead, an interface-like state appears, where half the pore is 
filled with an $\alpha$-like layer, and the rest by $\beta$-like phase. 
The position of the interface between both layers is controlled by the 
deviation of chemical potential with respect to the coexistence value, being 
possible to go from an $\alpha$-like capillary pure phase to a $\beta$-like 
one without undergoing any phase transition. The phase transition is only 
recovered in the bulk limit $H=\infty$. When the opposing surface
fields are not symmetrical (at least for near-symmetrical), it is 
expected to observe a similar behavior \cite{Parry}.

This analysis allows us to envisage different scenarios for the confined case 
and large $H$. For simplicity, we will only consider the case 
$\epsilon_W^1=\epsilon_W^2\equiv \epsilon_W$ in Eqs. (\ref{homeotropic})
and (\ref{planar}). We must note that this choice \emph{does not}
correspond to the symmetrical opposing fields, since the effect of each 
substrate is qualitatively different. For large $H$ and $\epsilon_W/\epsilon_A
<0.48$ and $\epsilon_W/\epsilon_A>0.53$ we expect to have a nematic-isotropic
capillary transition. The transition point will shifted towards lower or higher 
chemical potentials than the corresponding to the bulk transition if
$\cos \theta_1+ \cos \theta_2$ is negative or positive, respectively.
This state-dependent capillary transition shift, which does not occur for
simple fluids, has been observed in our model under confinement between 
symmetrical substrates that favor homeotropic anchoring \cite{paper2},
and also in confined spherocylinder model with the Onsager approximation
for a symmetrical slit pore \cite{Heras}. On the other hand, if 
$0.48<\epsilon_W/\epsilon_A<0.53$, no capillary transition is expected to 
be obtained for \emph{any} pore width. 

We present now our DFT calculations. For the sake of clarity, we separate
the results obtained for $\epsilon_W<\epsilon_W^{w1}$ and 
$\epsilon_W>\epsilon_W^d$.  

\subsection{The case $\epsilon_W<\epsilon_W^{w1}$}
%

We performed our first DFT calculations for $\epsilon_W/\epsilon_A=0.30$,
which correspond to partial wetting situation in the semi-infinite geometry 
for both substrates. Fig. \ref{fig4} shows the behavior of the averaged 
intrinsic order parameter $\bar U$ as the chemical potential is varied.
We see that there is a transition from an isotropic-like state 
(lower branch) to a nematic-like state (upper branch). The order of the 
transition is first-order, and it is found as the coincidence point of the 
grand-canonical free energies of the nematic-like and isotropic-like states.
We have shown also the metastable continuations of each branch: nematic-like
states on the left of the capillary NI transition and isotropic-like 
states on the right of the transition. Examination of the order parameter 
profiles show that the director field in the nematic-like state is oriented 
parallel throughout the pore to the substrate 
with an intrinsic order parameter similar to the bulk value, except in thin 
layers around the substrates, where the density and order parameter profiles 
are similar to the ones corresponding to the semi-infinite cases. 
In particular a very thin layer close to the $z=0$ substrate is 
homeotropically anchored. This state corresponds to the $S$ state 
in Ref. \onlinecite{paper3} and we will denote it by $N^S$. On the other hand,
the isotropic-like state (which we will denote as $I$) develops microscopic
homeotropically and random-planar layers close to the $z=0$ and $z=H$
substrates, respectively, and the orientational ordering decays to zero 
in the middle region of the pore. We have also compared our DFT calculations 
with the Kelvin law prediction Eq. (\ref{kelvin}) (see inset in Fig. 
\ref{fig4}). Although the transition value departs from this prediction
for small pore widths, for $H/d>20$ the agreement between our numerical 
estimations and the Kelvin predicted value is quite reasonable. 
Note that the capillary NI transition is shifted to larger chemical potentials
than the bulk NI transition value. However, as $\epsilon_W$ increases, this
trend is reversed (the Kelvin law predicts the turning point to occur at
$\epsilon_W/\epsilon_A=0.329$).

As $\epsilon_W$ exceeds $\epsilon_W^{w2}$, wetting in the $z=H$ wall begins
to affect the pore adsorption. For small values of $H$ a single transition
between an isotropic-like $I$ state to a nematic-like state occurs as for
$\epsilon_W<\epsilon_W^{w2}$. However, as $H$ increases the order of
the ``nematic-like'' state decreases and the chemical potential for which
this transition occurs converges to the $I-I^N$ transition value for the
$z=H$ wall. The order-parameter profiles make clear that the 
``nematic-like'' state corresponds to the formation of a nematic layer 
close to the $z=H$ substrate followed by an isotropic ordering along the
pore. By analogy with the semi-infinite situation, we call $I^N$ to the 
confined state after this transition, which we can see as a reminiscence
of the $I-I^N$ transition in the semi-infinite phenomenology. 
%

As $H$ increases, in addition to the $I-I^N$ transition, a new phase transition
between the $I^N$ to a proper nematic-like state $N^S$ (i.e. planar-anchored
along the pore) is observed. This transition can be regarded as the true
capillary NI transition. Fig. \ref{fig5} shows our DFT calculations for 
$H=20d$ and different values of $\epsilon_W$. It becomes clear that
the difference between the $I^N$ and the $N^S$ states 
at the transition decreases as $\epsilon_W$ increases, and eventually the
NI capillary transition disappears in a critical point (see inset of
Fig. \ref{fig5}). The value of $\epsilon_W$ at which the NI capillary
transition disappears is \emph{lower} than $\epsilon_W^{w1}$, i.e. the
confinement \emph{distabilizes} the two-phase region with respect to the 
one-phase region. We have also studied the effect of the pore width in the NI 
capillary transition. We set $\epsilon_W/\epsilon_A=0.465$, slightly below 
$\epsilon_W^{w1}$. Fig. \ref{fig6} shows that the capillary NI transition
does not exist for the smallest values of $H$ and reappears for $H>50d$. 
%

The disappearance of the NI capillar transition shown by our model is 
not related to the critical points observed for \emph{small} values of $H$ 
in other models \cite{Evans00,Heras}. Actually, those cases correspond to 
symmetrical slit-pores and consequently their phenomenology is not related 
to wetting properties of the substrates. In our model the NI capillary 
transition occurs at least for $H>20d$ when the substrates at $z=0$ and 
$z=H$ are the same \cite{paper2}. However we cannot rule out the possibility 
of critical points for very small $H$, but we will not consider them since 
the predictions of our model are not reliable in that regime.

Our DFT results about the existence of the NI capillary transition 
show the opposite trend to the one predicted by a Landau-de Gennes model 
in Ref. \onlinecite{Robledo98}, which predicted that the confinement
\emph{stabilized} the two-phase region with respect to the one-phase region.
Moreover for large values of $H$ the boundary for the non-existence
region of the capillary transition was a triple point rather than a critical 
point. Actually the scenario shown by our DFT calculations is consistent
with the phenomenology that presents a confined Ising model with opposing
surface fields when the wetting transition is critical \cite{Parry}.
This analogy can be explained heuristically. For large $H$ and $\epsilon_W$ 
close to $\epsilon_W^{w1}$, the typical configuration close to bulk 
coexistence corresponds to an interfacial state formed by an isotropic layer 
around the $z=0$ and a nematic layer around the $z=H$ wall. In first 
approximation the NI interface position is given by the minimum binding 
potential which is the superposition of the contributions from the two 
semi-infinite systems \cite{Parry}. As there is a complete wetting 
situation at the $z=H$ substrate, we can consider its binding potential 
as a exponentially decreasing function of the interfacial position with 
respect to the $z=H$ wall $c\exp[-(H-l)/\xi_b']$, where $c>0$ and 
$\xi_b'$ is the nematic phase correlation length at bulk NI coexistence. 
Note that the local minimum corresponding to the bounded state, if any, 
plays no role in our argument as soon as we are far enough from the 
first-order wetting transition corresponding to the semi-infinite case with 
the $z=H$ wall. On the other hand, the binding potential corresponding to the
$z=0$ substrate takes the usual form $-a\exp(-l/\xi_b)+b\exp(-2l/\xi_b)$, where
$a\propto \epsilon_W^{w1}-\epsilon_W >0$, $b>0$ and $\xi_b$ is the
isotropic phase correlation length at bulk NI coexistence. So, the complete 
binding potential for the NI interfase, up to irrelevant constants, is:
\begin{equation}
W(l)=h(H-l)-ae^{-\frac{l}{\xi_b}}+be^{-\frac{2l}{\xi_b}}+
ce^{-\frac{H-l}{\xi_b'}}
\label{binding}
\end{equation}
where $h\propto \Delta \mu$. A critical point must verify:
\begin{eqnarray}
h&=&\frac{a}{\xi_b}e^{-\frac{l}{\xi_b}}-\frac{2b}{\xi_b}e^{-\frac{2l}{\xi_b}}
+\frac{c}{\xi_b'} e^{-\frac{H-l}{\xi_b'}}\label{m1}\\
0&=&-\frac{a}{\xi_b^2}e^{-\frac{l}{\xi_b}}+\frac{4b}{\xi_b^2}e^{-\frac{2l}
{\xi_b}}+\frac{c}{(\xi_b')^2} e^{-\frac{H-l}{\xi_b'}}\label{m2}
\end{eqnarray}
Instead of solving simultaneously Eqs. (\ref{m1}) and (\ref{m2}), we 
will study the spinodal line, which is the solution of Eq. (\ref{m2}). 
We consider solutions of the form $l\equiv \chi H + \Delta l$, where $0< \chi
< 1$ and $\Delta l\sim \xi_b,\xi_b'$. Substituting this ansatz in Eq. 
(\ref{m2}), we obtain
\begin{eqnarray}
0&=&-\left(ae^{-\frac{\chi H}{\xi_b}}\right)e^{-\frac{\Delta l}{\xi_b}}
+\left(4be^{-\frac{2\chi H}{\xi_b}}\right)e^{-\frac{2\Delta l}{\xi_b}}
\nonumber\\ &+&\left(c\frac{\xi_b^2}{(\xi_b')^2}e^{-\frac{(1-\chi)H}
{\xi_b'}}\right) e^{\frac{\Delta l}{\xi_b'}}
\label{m2-2}
\end{eqnarray}
In order to have two solutions for the Eq. (\ref{m2-2}), we need all
the factors in brackets in Eq. (\ref{m2-2}) to be of the same order. 
Consequently we obtain that $\chi=\xi_b/(2\xi_b'+\xi_b)$ and 
$a=a'\exp[-H/(2\xi_b' +\xi_b)]$, where $a'\sim b\sim c$. The value of 
$a'$ for which the critical point occurs corresponds to the case where the 
rescaled equation:
\begin{equation}
0=-a'e^{-\frac{\Delta l}{\xi_b}}+4be^{-\frac{2\Delta l}{\xi_b}}
+c\frac{\xi_b^2}{(\xi_b')^2}e^{\frac{\Delta l}{\xi_b}}
\label{m2-3}
\end{equation}
has an unique real solution. For simplicity, we assume that $\xi_b=\xi_b'$, 
but similar results are found when both correlation lengths are different. 
We refrain to give explicit solutions and we just mention that indeed 
$a'/c \sim 1$ and the value of $\Delta l$ at the critical point is
$\Delta l_c \sim \xi_b$ for reasonable values of $b$ and $c$. 
Consequently, the interfacial position at the critical point
$l_c\sim H/3$ and the NI capillary transition shift 
$\epsilon_W^{w1}-\epsilon_W\sim \exp(-H/3\xi_b)$ and from Eq. (\ref{m1}) 
$\Delta \mu \sim \exp(-2H/3\xi_b)$. 

A numerical comparison of these predictions with our DFT calculations, 
although it would be very interesting, is extremely difficult due to the 
limitations of our numerical procedure and it will not be carried out
in this work. However, at least qualitatively these predictions have been
confirmed by the DFT calculations even for not so large values of $H$.

\subsection{The case $\epsilon_W>\epsilon_W^d$}
%

Now we turn to study the confined system for $\epsilon_W>\epsilon_W^d$.
We first considered the case $\epsilon_W/\epsilon_A=0.70$ and
$H=20d$. Different confined states are obtained as the chemical potential 
increases (see Fig. \ref{fig7}). The first cases Figs. \ref{fig7}(a) and 
\ref{fig7}(b) clearly correspond to a vapor-like and isotropic-like states, 
and we will denote them as $V$ and $I$, respectively. There is in both cases 
a homeotropically-anchored nematic layer close to the substrate at $z=0$ and 
a random planar layer close to the substrate at $z=H$. As the chemical 
potential increases, the random planar layer at $z=H$ undergoes a transition
to a planar biaxial state. Again this first-order transition is a reminiscence
of the $I-I^N$ surface transition corresponding to the substrate at $z=H$,
and we will denote both confined states $I$ and $I^N$ by analogy with the
semi-infinite phenomenology. As the chemical potential
is further increased, the nematic layer increases and a state in which an
isotropic and a nematic layer coexist is obtained (Fig. \ref{fig7}(c)). 
There is no phase transition between the isotropic-like $I^N$ and 
the nematic-like $N^S$ states, but a smooth transformation similar to the 
expected one when both substrates are completely wet by different bulk phases 
at coexistence. Finally, there is a transition to a new nematic-like state 
(the \emph{L state} in Ref. \onlinecite{paper3}, that we will denote as $N^L$), 
characterized by an almost linear distorsion from homeotropic to planar 
configuration on the tilt angle through the pore, although the 
\emph{intrinsic} order parameter takes essentially the bulk value except 
close to the substrates (see Fig. \ref{fig7}(d)). This transition between 
the $N^L$ and $N^S$ state is first order and converges to the anchoring 
transition that is observed in the semi-infinite case with a substrate that 
favors a homeotropic anchoring as $H\to \infty$ \cite{paper1}. A more 
detailed discussion about this transition can be found in 
Ref. \onlinecite{paper3}. 
%

For larger values of $H$, the capillary NI transition appears as a transition
between an $I^N$ state and a $N^L$ state for chemical potentials smaller
than the bulk NI transition value. As $\epsilon_W$ approaches $\epsilon_W^d$,
the range of values of $H$ for which capillary transition does not exist 
is wider. For $\epsilon_W/\epsilon_A=0.53$ (close to the dewetting 
transition point) the capillary NI transition does not appear until
$H/d>150$. Fig. \ref{fig8} shows the continuous change of the nematic layer
as the chemical potential increases for $H=20d$, which is analogous to the 
behavior observed for $\epsilon_W/\epsilon_A=0.70$. Fig. \ref{fig9} shows 
how the curves of the averaged intrinsic order parameter as a function of the 
chemical potential change for increasing pore widths. For the narrowest pore 
($H=10d$), a single transition is observed between an isotropic state to an 
state with some nematic ordering.
However, as $H$ is increased, it becomes clear that this transition is again a 
reminiscence of the $I-I^N$ transition in the semi-infinite case. The 
localization of this transition depends weakly on the pore width $H$. 
For $H>13d$, the $N^L-N^S$ transition appears at chemical potentials larger 
than the bulk NI coexistence value. This transition does not appear for 
$H=10d$, which is a clear indication of the existence of a critical point 
for the $N^L-N^S$ transition at small values of $H$. As $H$ increases, the 
chemical potential value for $N^L-N^S$ transition shifts to lower values.
Finally, we observe that the $\bar U$ curve becomes increasingly steep at
the bulk NI transition as $H\to \infty$. This fact is a 
consequence of the formation of interfacial-like states (as Fig. 
\ref{fig7}(c)) in the proximities of the bulk NI coexistence chemical 
potential.
%

The non-existence of a capillary NI transition in the situation described 
above is unexpected. First, the finite-width effects are much more pronounced
for $\epsilon_W>\epsilon_W^d$ than below $\epsilon_W^{w1}$. As a consequence,
we observe significant departures from the bulk behavior for wider pores and
in a larger range of $\epsilon_W$ than when we approach the complete wetting
situation from below (i.e. $\epsilon_W<\epsilon_W^{w1}$). This fact
suggests the emergence of a new length scale in the system.
On the other hand, as the dewetting transition is first order, 
we could naively expect that under confinement and large enough $H$ the 
capillary transition disappearance point is shifted to \emph{lower} values of
$\epsilon_W$ \cite{Indekeu,Robledo98} (recall that the role of $\epsilon_W$ or
the temperature is reversed in the dewetting transition with respect to 
the wetting transition). To understand this situation we recall that the
confinement in a hybrid cell distabilizes the $N^L$ with respect to the
$N^S$. A simple phenomenological argument can be used to explain this fact
\cite{paper3}. For large enough values of $H$, the excess free-energy per
surface area of a $N^S$ state is given by the sum of the surface free energies 
between each substrate and a nematic fluid which is planar-anchored with 
respect to them. The finite-width corrections are expected to decay 
exponentially. On the other hand, the large $H$ behavior free energy of the 
$N^L$ state is the sum of the surface tensions of the interface between the 
$z=0$ substrate and a homeotropically-anchored nematic fluid and the 
corresponding between the planar-anchored nematic fluid and the $z=H$ 
substrate. For this state the finite-width correction is dominated by the 
elastic deformations contribution and decays as $K_3\pi^2/8 H$ as 
$H\to \infty$, where $K_3$ is a Frank-Oseen elastic constant \cite{paper3}.
So, for a given temperature, the $N^S-N^L$ transition is always shifted
towards higher chemical potentials than the anchoring transition, and the
convergence to it for large values of $H$ is quite slow. Although this argument
implicitely assumes that the inhomogeneities are restricted to microscopic
regions around each substrate, it is also valid if we allow the $N^S$ state
to develop an isotropic layer close to the $z=0$ substrate (as occurs close
to the bulk NI coexistence). Consequently, for $\epsilon_W^d<\epsilon_W <
\epsilon_W^m$ and moderate values of $H$, the most stable nematic-like state 
close to the bulk NI coexistence is $N^S$, and therefore there is no
capillary transition. As $H$ increases, the $N^L$ state becomes more stable 
than the $N^S$ state near the $NI$ bulk transition, and eventually the 
capillary transition between a $I^N$ state and a $N^L$ occurs.

An interesting feature of the NI capillary transition in this regime is
that it is linked to the \emph{purely surface transition} $N^S-N^L$. 
As explained before, for small values of $H$ there is no NI capillary 
transition but if $H$ is not too small there is a $N^S-N^L$ transition.
As $H$ increases, the chemical potential corresponding to the
$N^S-N^L$ transition shifts to smaller values, and eventually crosses the
value corresponding to the bulk NI transition (see Fig. \ref{fig10}). 
When this happens, the transition is between a isotropic-like $I^N$ state 
to the nematic-like $N^L$ state: the capillary NI transition is recovered. 
Finally, the Kelvin relationship Eq. (\ref{kelvin}) predicts that the 
capillary transition value of $\beta \mu$ approaches the bulk NI value from 
\emph{below}. The DFT results for large $H$ support this prediction. So, at 
fixed temperature and chemical potential, and changing $H$, reentrance of the 
isotropic-like state is observed (see Fig. \ref{fig10}). 
%

\section{Conclusions}

In this paper we have reported a DFT study of the NI capillary transition
under confinement in a slit-pore where each substrate favors a different
anchoring condition. We found that there is an interplay between the 
phenomenology found in the semi-infinite problems with the phenomenology 
which appeared in the confined problem. We have focussed in the non-existence
of the NI capillary transition as a result of both substrates to be wet
by different phases. If we approach this region from below, we observe
a distabilization of the two-phase region with respect to the one-phase region.
Moreover, the capillary NI transition finishes in a critical point. These
findings are different from the ones reported by Quintana \emph{et al} 
\cite{Robledo98} and this discrepancy can be traced to the different order of 
the wetting transition. When we approach the non-existence range for the
NI capillary transition from above, a different mechanism is involved in 
the finite-size effects of the capillary NI transition. Although the wetting 
transition which drives the disappearance of the capillary transition is 
first-order, the two-phase region is still distabilized with respect to the 
one-phase region. The explanation for this is the existence of two 
nematic-like locally stable phases with a relative stability dependent on the 
pore width. One of them corresponds to a complete wetting situation, so 
links with the isotropic-like states without undergoing a phase transition. 
So, as $H$ increases, the phase transition between two nematic-like states 
transforms continuously to the NI capillary transition. As a consequence, a 
reentrant behavior for the isotropic-like state is observed by varying the 
pore width and keeping the temperature and chemical potential constant. 

Although we have restricted our study to an isotherm case, we anticipate that
we would find a similar behavior if the temperature is changed instead of
the value of $\epsilon_W$. On the other hand, the condition $\epsilon_W^1=
\epsilon_W^2$ is not so restrictive as soon as the wetting transition 
of the $z=H$ wall precedes the complete wetting at the $z=0$ wall. If such
a condition is not fulfilled, our conclusions may change. 

\section*{Acknowledgments}
I.R.P. would like to thank Prof. F. Schwabl for his help in the development 
of this work and Prof. J. O. Indekeu for useful discussions. J. M. R.-E.
also thanks Prof. A. O. Parry for his interest in this work. I.R.P note that 
this work has been performed under the auspices of the Sonderforschungsbereich 
(SFB) 563. Financial support by the Deutsche Forschungsgemainschaft (DPG) 
is gratefully acknowledged.  J.M. R.-E. and L.F.R acknowledge finantial 
support by Grant No. BQU2001-3615-C02-02 from MEC (Spain). Finally, 
J. M. R.-E. also acknowledges partial financial support from 
Secretar\'{\i}a de Estado de Educaci\'on y Universidades (Spain), 
co-financed by the European Social Fund, and from 
the European Commission under Contract MEIF-CT-2003-501042.

\clearpage
\begin{figure}
\caption{The adsorption $\Gamma$ plotted vs. the chemical potential for a
semi-infinite geometry where the fluid is isotropic in bulk and the substrate
favors homeotropic anchoring. The dashed line corresponds to 
$\epsilon_W/\epsilon_A=0.3$ and the continuous line to 
$\epsilon_W/\epsilon_A=0.66$. The $I-I^N$ transition for the latter case
is represented by the dotted line. The insets represent the reduced
density $\rho^*\equiv \rho d^3$ and extrinsic order parameter profiles 
$\eta$, $\sigma$ and $\nu$ for different chemical potentials and 
$\epsilon_W/\epsilon_A=0.66$.\label{fig1}}
\end{figure}
\begin{figure}
\caption{Plot of the reduced density $\rho d^3$ 
and extrinsic order parameter profiles $\eta$, $\sigma$ and $\nu$ for 
$\epsilon_W/\epsilon_A=0.70$ and $\beta \mu=-3.918$, obtained by numerical 
minimization for $R=435d$. Inset: the intrinsic order 
parameters $U$ and $\psi$ corresponding to the same state. The intrinsic order
parameter $B$ is zero except close to $z=R$ (not shown for clarity).
\label{fig2}}
\end{figure}
\begin{figure}
\caption{Plot of the cosine of the contact angle $\theta$ of an isotropic
phase droplet on the substrate-nematic interface as a function of the 
fluid-substrate interaction strength $\epsilon_W$ (thick continuous line). 
The filled circles correspond to the numerical estimates for $\epsilon_W>
\epsilon_W^{w3}$ (see text for explanation).
\label{fig3}}
\end{figure}
\begin{figure}
\caption{Plot of the averaged intrinsic order parameter throughout the pore
$\bar U$ as a function of the reduced chemical potential $\beta \mu$ 
for $\epsilon_W/\epsilon_A=0.3$ and $H/d=10$ (continuous line) and 
$H/d=20$ (dashed line). The dotted lines show the location of the capillary
NI transition. Inset: the diamonds represent our DFT values of the chemical 
potential for $\epsilon_W/\epsilon_A=0.3$ at the capillary NI transition 
for $H/d=10$, $20$ and $35$, and the continuous line is the Kelvin law 
prediction, Eq. (\ref{kelvin}).
\label{fig4}}
\end{figure}
\begin{figure}
\caption{Plot of the averaged intrinsic order parameter throughout the pore
$\bar U$ as a function of the reduced chemical potential $\beta \mu$ 
for $H/d=20$ and $\epsilon_W/\epsilon_A=0.35$ (dashed line), $\epsilon_W/
\epsilon_A=0.41$ (continuous line) and $\epsilon_W/\epsilon_A=0.45$ 
(dot-dashed line). The dotted lines show the location of the $I-I^N$ 
transition (left) and capillary NI transitions (right). Inset: The symbols are
the coexistence values of $\bar U$ at the $I-I^N$ and capillary NI transitions
for different values of $\epsilon_W$. The continuous lines are to guide the 
eye.\label{fig5}}
\end{figure}
\begin{figure}
\caption{Plot of the averaged intrinsic order parameter throughout the pore
$\bar U$ as a function of the reduced chemical potential $\beta \mu$ 
for $\epsilon_W/\epsilon_A=0.465$ and $H/d=10$, $20$, $40$, $50$ and $60$.
\label{fig6}}
\end{figure}
\begin{figure}
\caption{Plot of the density ($\rho^*\equiv\rho d^3$) 
and intrinsic uniaxial order parameter
($U$) for the fluid confined in a pore of width $H=20d$ and $\epsilon_W/
\epsilon_A=0.70$ and chemical potentials: (a) $\beta \mu=-4.7$, (b) $\beta 
\mu=-4.45$, (c) $\beta \mu=-3.95$ and $\beta \mu=-3.7$. 
\label{fig7}}
\end{figure}
\begin{figure}
\caption{Density ($\rho^*\equiv \rho d^3$) and intrinsic uniaxial order 
parameter profiles ($U$) for $H=20d$ and $\epsilon_W/\epsilon_A=0.53$. 
The chemical potentials run from $\beta \mu=-4.03$ to $\beta \mu=-3.64$ with 
a step of $\Delta(\beta\mu)=0.01$.
\label{fig8}}
\end{figure}
\begin{figure}
\caption{Plot of the averaged intrinsic order parameter $\bar U$
throughout the pore vs the reduced chemical potential $\beta \mu$ 
for $\epsilon_W/\epsilon_A=0.53$ and $H/d=10$, $13$, $20$, $30$, $40$
and $80$. The discontinuities of $\bar U (\beta \mu)$ correspond
to the $I-I^N$ (left) and $N^S-N^L$ (right) transitions. \label{fig9}}
\end{figure}
\begin{figure}
\caption{Plot of the reduced chemical potential $\beta\mu$ corresponding 
to the $N^S-N^L$ transition (or the $NI$ capillary transition, see text for 
explanation) as a function of the reduced inverse pore width $d/H$ for values 
of $\epsilon_W/\epsilon_A=0.49$ (circles), $0.53$ (diamonds), $0.60$ 
(triangles up) and $0.66$ (triangles down). The continuous lines are only to
guide the eye.
\label{fig10}}
\end{figure}
\clearpage
\begin{figure}
\includegraphics[width=15cm]{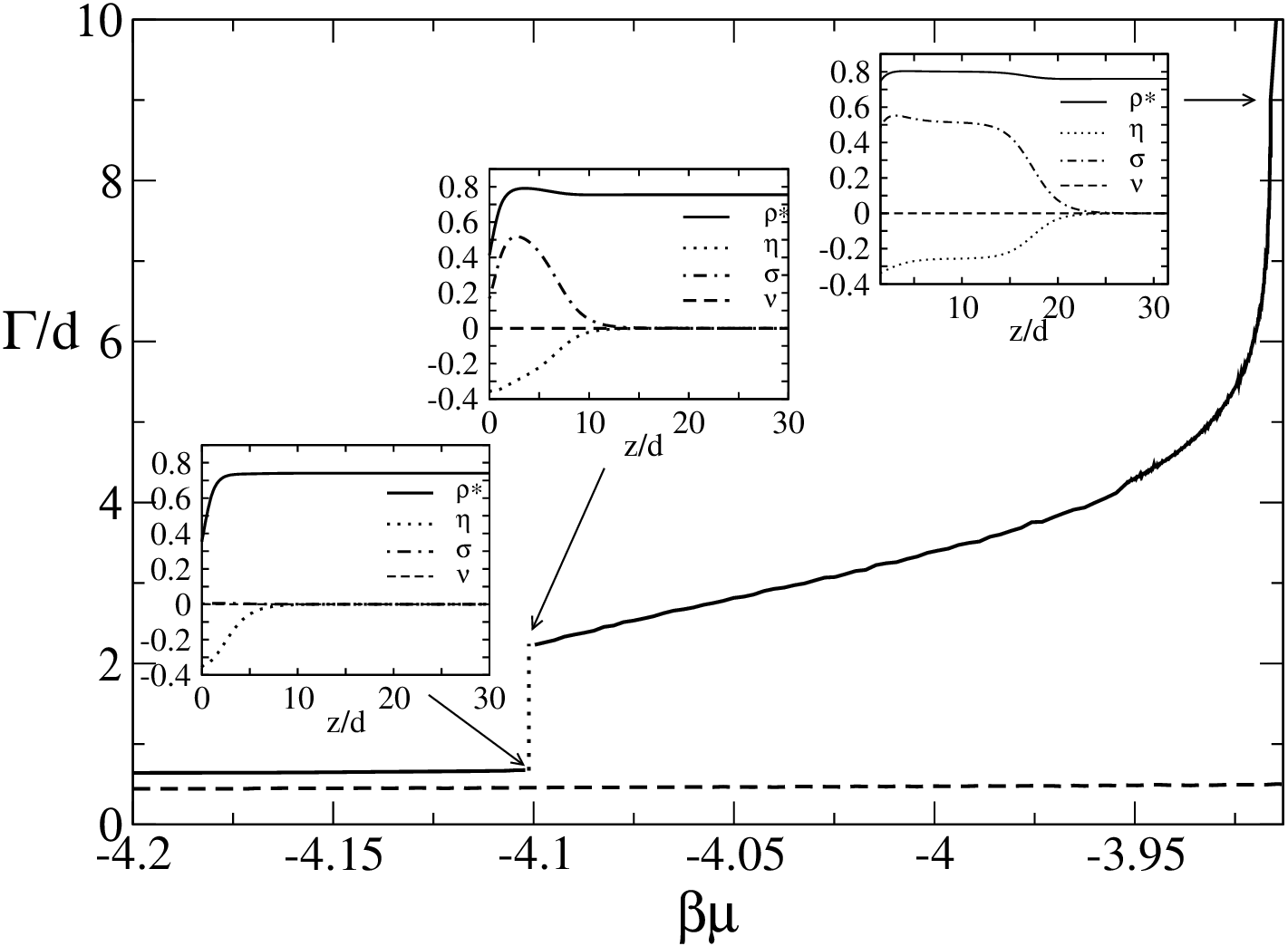}
\end{figure}
\vspace*{3cm}
\hspace{2.5cm}
{\Large I. RODRIGUEZ-PONCE \emph{et al}, FIGURE 1}
\clearpage
\begin{figure}
\includegraphics[width=15cm]{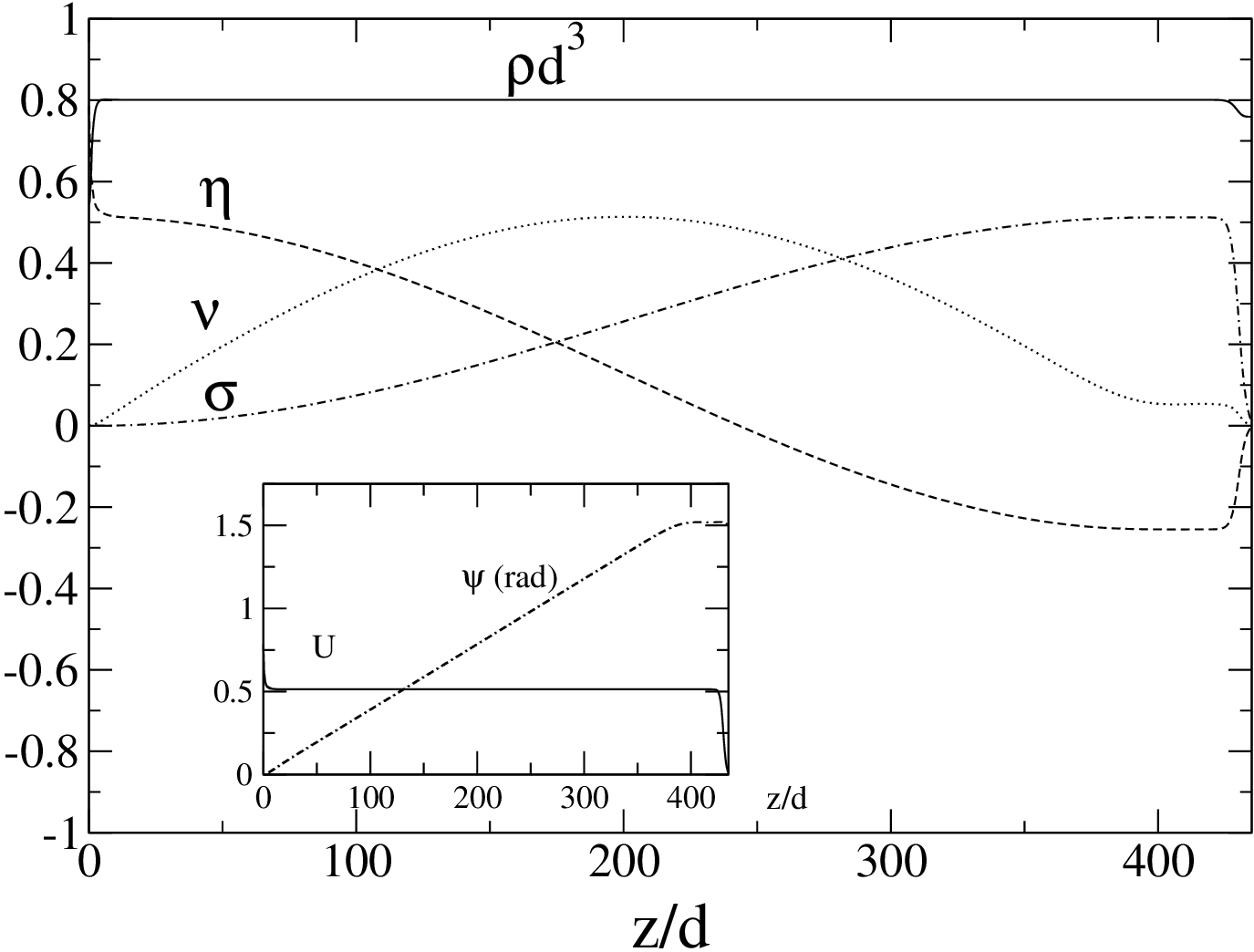}
\end{figure}
\vspace*{3cm}
\hspace{2.5cm}
{\Large I. RODRIGUEZ-PONCE \emph{et al}, FIGURE 2}
\clearpage
\begin{figure}
\includegraphics[width=15cm]{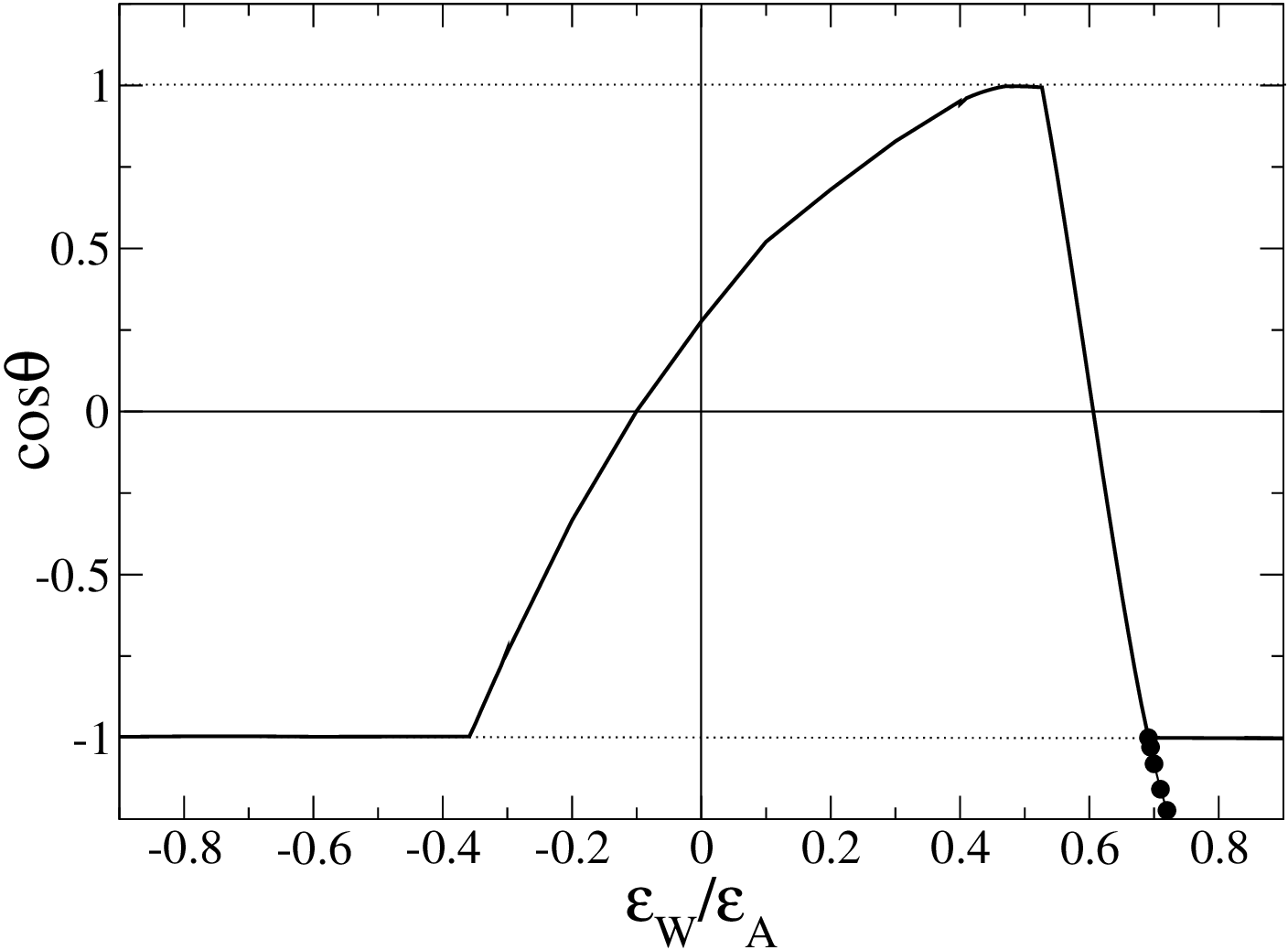}
\end{figure}
\vspace*{3cm}
\hspace{2.5cm}
{\Large I. RODRIGUEZ-PONCE \emph{et al}, FIGURE 3}
\clearpage
\begin{figure}
\includegraphics[width=15cm]{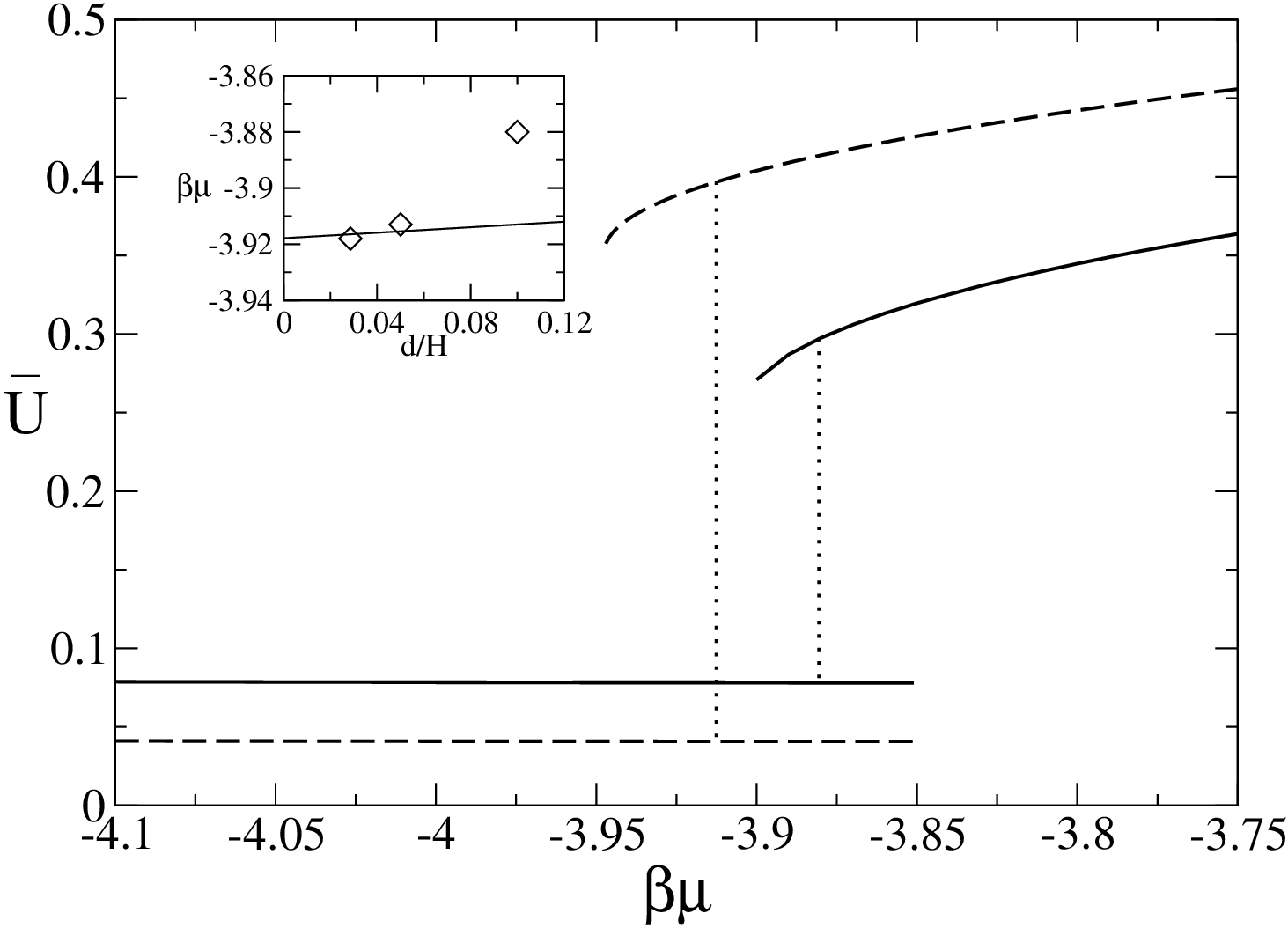}
\end{figure}
\vspace*{3cm}
\hspace{2.5cm}
{\Large I. RODRIGUEZ-PONCE \emph{et al}, FIGURE 4}
\clearpage
\begin{figure}
\includegraphics[width=15cm]{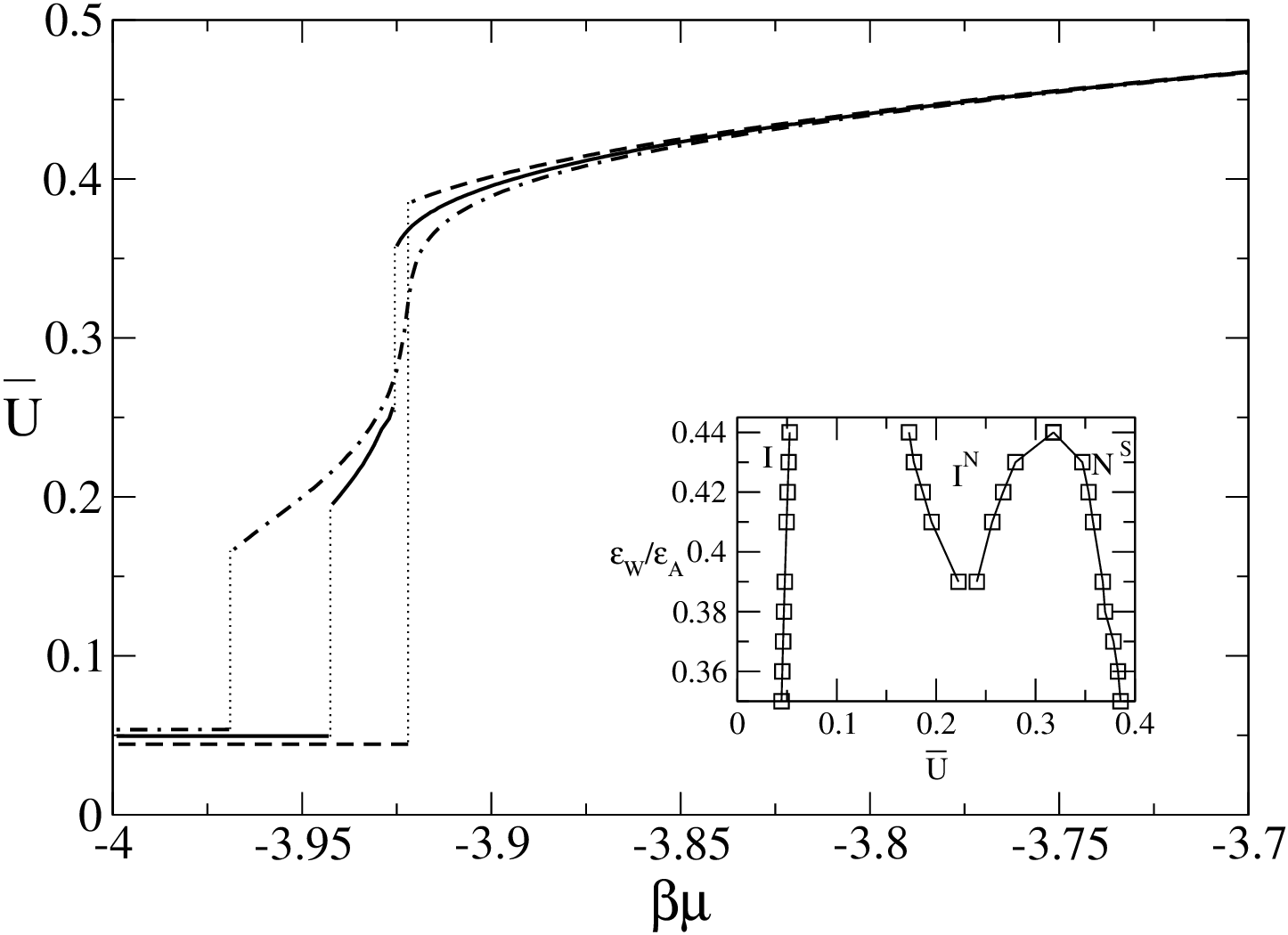}
\end{figure}
\vspace*{3cm}
\hspace{2.5cm}
{\Large I. RODRIGUEZ-PONCE \emph{et al}, FIGURE 5}
\clearpage
\begin{figure}
\includegraphics[width=15cm]{fig6.eps}
\end{figure}
\vspace*{3cm}
\hspace{2.5cm}
{\Large I. RODRIGUEZ-PONCE \emph{et al}, FIGURE 6}
\clearpage
\begin{figure}
\includegraphics[height=15cm,angle=270]{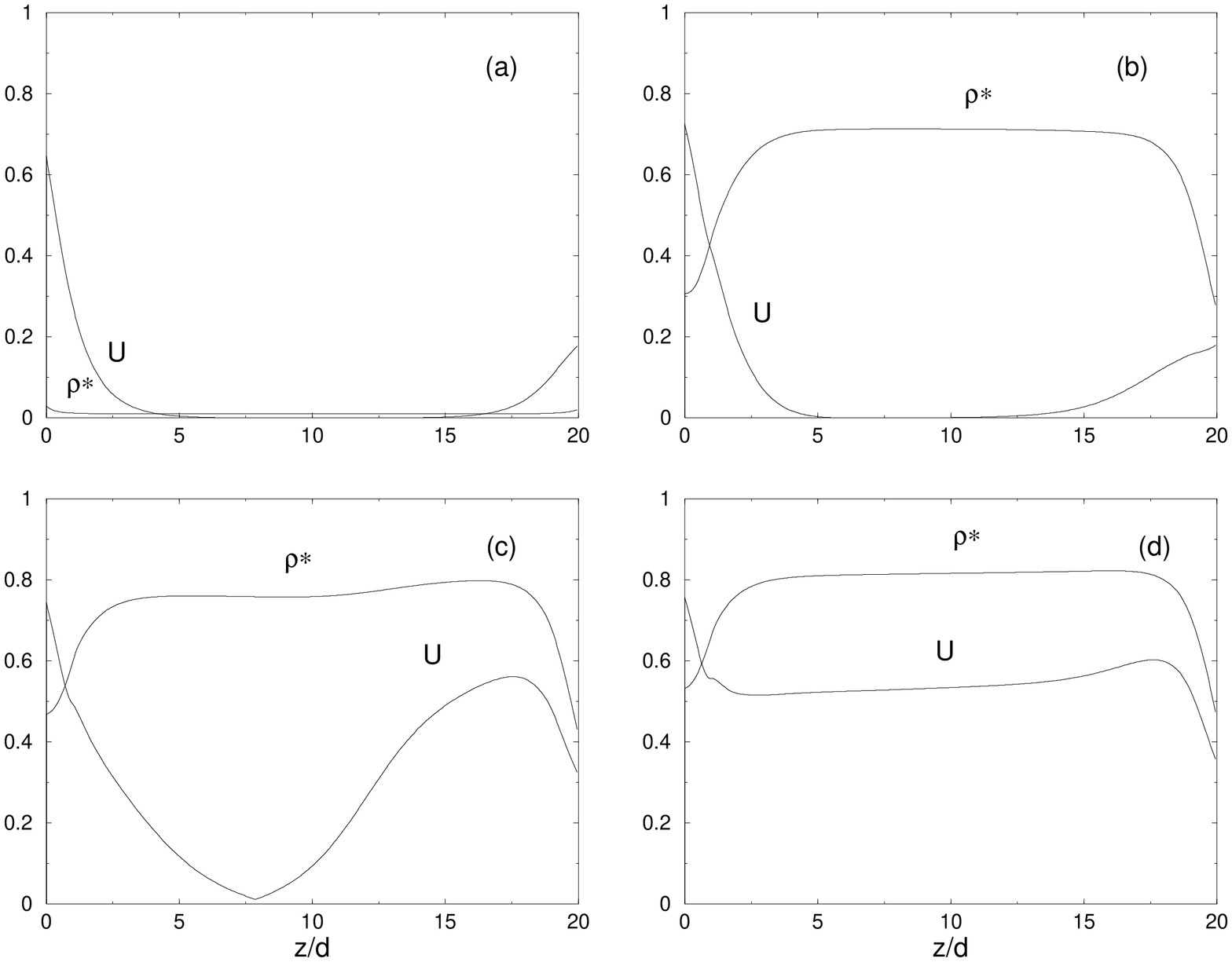}
\end{figure}
\vspace*{3cm}
\hspace{2.5cm}
{\Large I. RODRIGUEZ-PONCE \emph{et al}, FIGURE 7}
\clearpage
\begin{figure}
\includegraphics[height=15cm,angle=270]{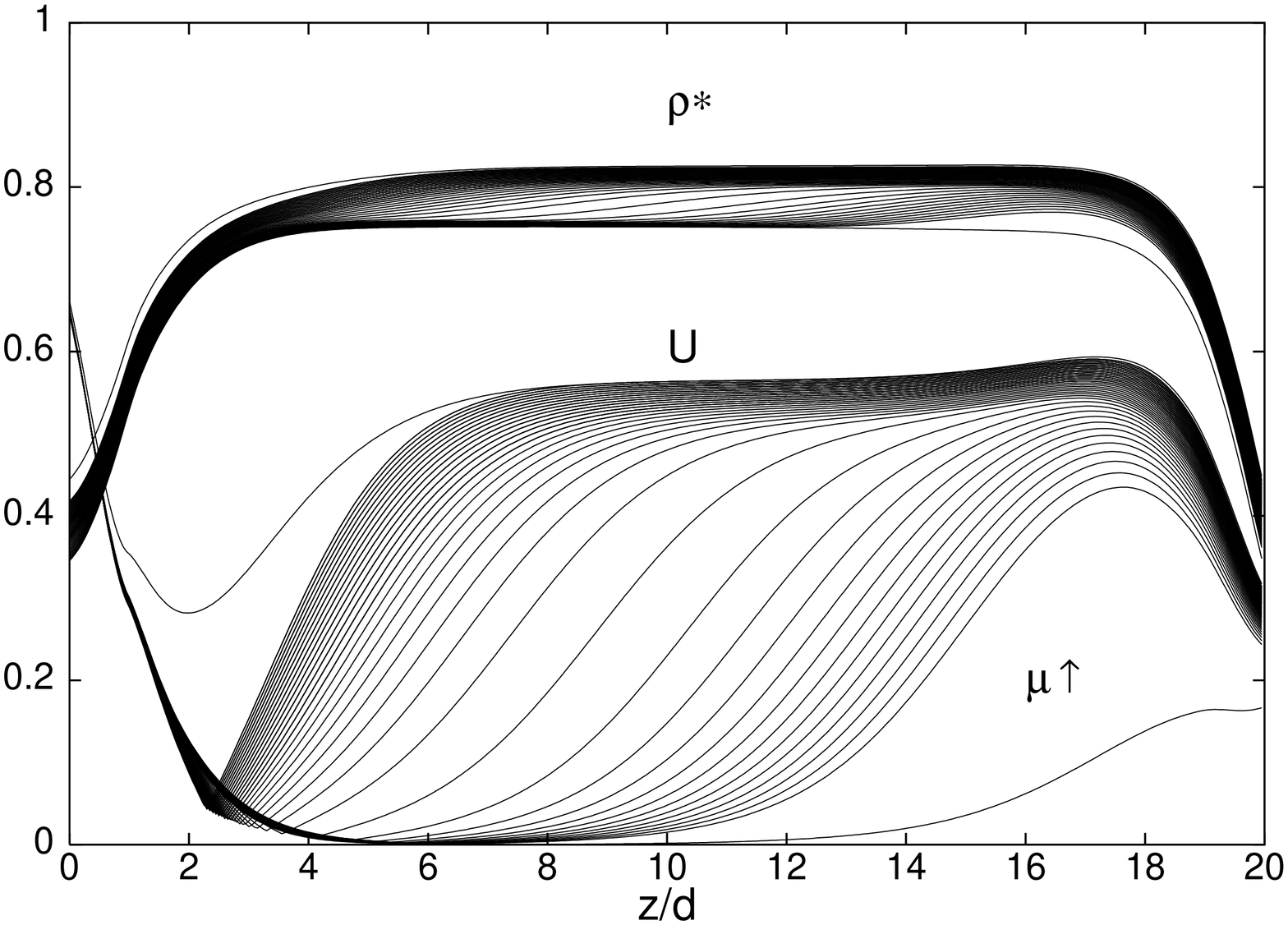}
\end{figure}
\vspace*{3cm}
\hspace{2.5cm}
{\Large I. RODRIGUEZ-PONCE \emph{et al}, FIGURE 8}
\clearpage
\begin{figure}
\includegraphics[height=15cm,angle=270]{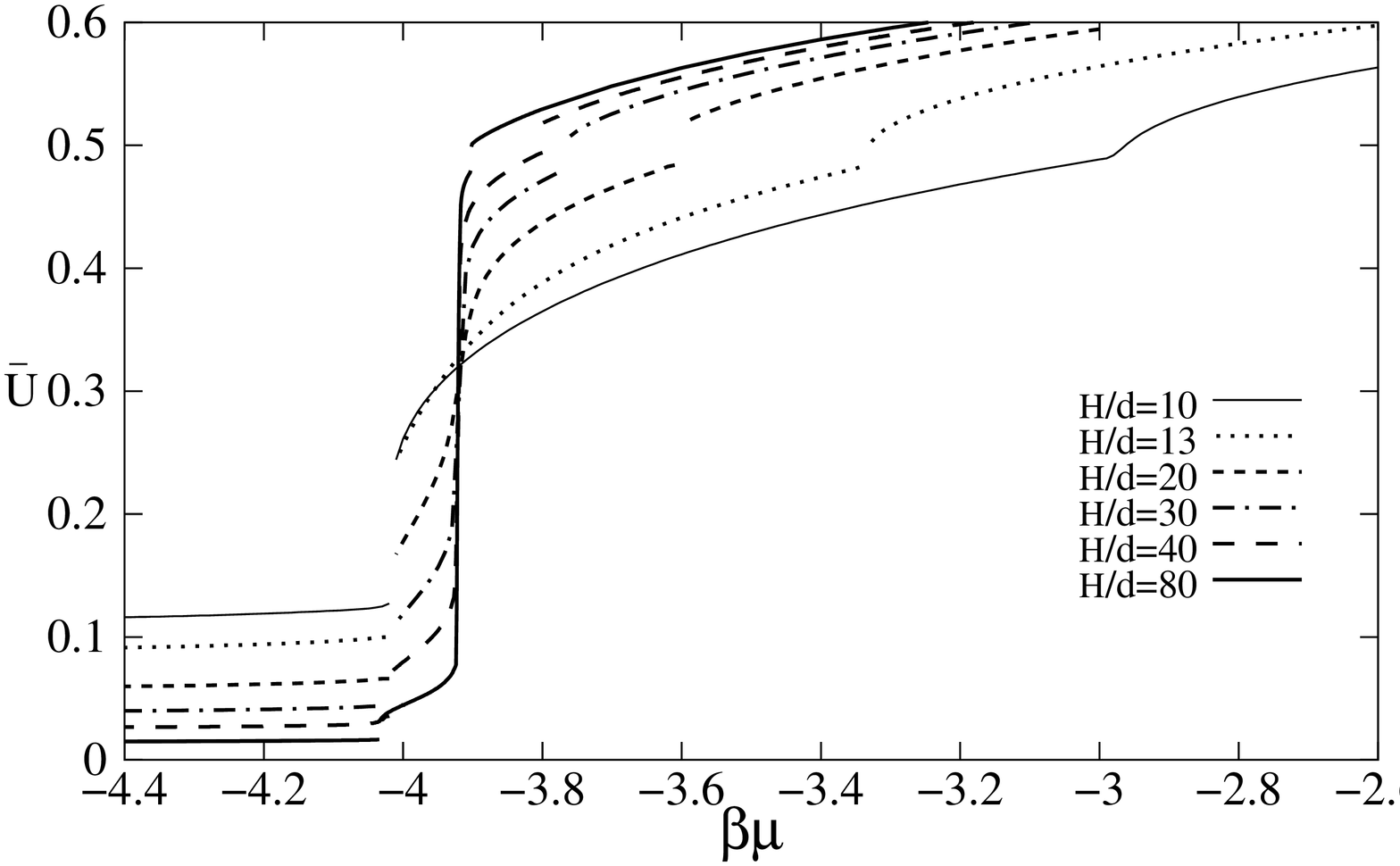}
\end{figure}
\vspace*{3cm}
\hspace{2.5cm}
{\Large I. RODRIGUEZ-PONCE \emph{et al}, FIGURE 9}
\clearpage
\begin{figure}
\includegraphics[width=15cm]{fig10.eps}
\end{figure}
\vspace*{3cm}
\hspace{2.5cm}
{\Large I. RODRIGUEZ-PONCE \emph{et al}, FIGURE 10}
\end{document}